\pdfoutput=1
\documentclass[aps,prc,floatfix,reprint,longbibliography,nofootinbib,showkeys]{revtex4-1}

\usepackage{amsmath}    
\usepackage{graphicx}   
\usepackage{color}
\usepackage{siunitx}  
\usepackage[colorlinks,linkcolor=blue,citecolor=blue]{hyperref}   
\usepackage{bm} 
\usepackage{import}

\newcommand\nda{\end{align}}
\def\Eq#1{Eq.~(\ref{#1})}
\def\app#1{Appendix~\ref{#1}}
\def\Fig#1{Fig.~\ref{#1}}

\def\bra{\left\langle}
\def\ket{\right\rangle}
\def\E{\mathcal{E}}
\def\atanh{\rm atanh}
\def\t{\tilde}

\advance\parskip 1.9pt
\advance\voffset -0.2in

\def\be{\begin{equation}}
\def\ee{\end{equation}}
\def\bg{\begin{eqnarray}}
\def\nd{\end{eqnarray}}

\begin{document}


\title{Pseudorapidity dependent hydrodynamic response in 
heavy-ion collisions}


\author{Hui Li}
\author{Li Yan}
\email[]{cliyan@fudan.edu.cn}
\affiliation{Key Laboratory of Nuclear Physics and Ion-Beam Application (MOE) \& Institute of Modern Physics\\
Fudan University, 220 Handan Road, 200433, Yangpu District, Shanghai, China}

\date{\today}

\begin{abstract}
We propose a differential hydrodynamic response relation,
$V_2(\zeta)=\int d\xi G(\zeta-\xi) \mathcal{E}_2(\xi)$, to describe the formation of a pseudorapidity 
dependent elliptic flow in heavy-ion collisions, 
in response to a fluctuating three-dimensional initial density profile. By analyzing the medium expansion
using event-by-event simulations of 3+1D MUSIC, with initial conditions generated via the AMPT model, 
the differential response relation is verified. Given the response relation, we are able to separate the two-point correlation of elliptic flow in 
pseudorapidity into fluid response  
and two-point correlation of initial eccentricity. The fluid response contains information
of the speed of sound and shear viscosity of the medium. From the pseudorapidity dependent 
response relation, a finite radius of convergence of the hydrodynamic gradient expansion is obtained 
with respect to  realistic fluids in heavy-ion collisions.  
\end{abstract}
\maketitle

\section{Introduction} 
In high energy heavy-ion experiments, the fluid nature of quark-gluon plasma (QGP) has been well demonstrated
through the measurements of flow harmonics in multi-particle correlations~\cite{Song:2017wtw}. 
These flow harmonics can be understood as 
the fluid response to the decomposed azimuthal modes associated with
the initial state geometrical deformations, giving rise to 
a series of response relations (cf.~\cite{Heinz:2013th,Yan:2017ivm} for recent reviews). In particular,
the eccentricity of the initial density profile $\E_2$, which characterizes the elliptic 
deformation of the QGP fireball at initial time,
and elliptic flow $V_2$~\cite{Ollitrault:1992bk}, which characterizes the asymmetric emission of final state particles in azimuthal angles, 
are linearly correlated: 
\be
\label{eq:linear}
V_2=G_0 \E_2\,.
\ee
Both $V_2$ and $\E_2$ fluctuate from event to event, but 
this linear relation has been found valid for the mid-central collisions via 
event-by-event hydrodynamic simulations~\cite{Qiu:2011iv,Niemi:2012aj,Noronha-Hostler:2015dbi}.
In \Eq{eq:linear}, the response coefficient $G_0$ depends on the fluid properties of QGP, which is a real constant 
in one specified centrality class. 
A consistent suppression of the response coefficient has been observed when dissipative 
corrections in the QGP become larger.

\Eq{eq:linear} 
has lead to many remarkable applications 
in heavy-ion collisions, such as
the background subtraction of chiral magnetic effects~\cite{Kharzeev:2007jp}
using a selected collision geometry~\cite{Skokov:2016yrj,Voloshin:2010ut}.
Nevertheless, as one relation between the \emph{global} azimuthal asymmetry of initial and final states,
knowledge of the \emph{local} structure of the QGP medium is absent.
In this letter, we propose a generalization of \Eq{eq:linear} to a pseudorapidity dependent
hydrodynamic response. 
Our generalization is partly motivated by the experimental developments in heavy-ion collisions, where pseudorapidity dependent 
flow harmonics have been 
explored~\cite{Khachatryan:2015oea,Aad:2014eoa,Adam:2016ows}. 
Besides,  the pseudorapidity dependent hydrodynamic response 
is expected to play a more significant role in the beam energy scan program carried out at the Relativistic
Heavy-Ion Collider~\cite{Luo:2017faz}, since at lower collision energies, 
approximations for the
fluid dynamics along the longitudinal direction
with respect to Bjorken symmetry are no longer valid 
in theoretical models~\cite{Du:2019obx}.

A pseudorapidity dependent generalization of \Eq{eq:linear} reveals
the local properties of the QGP medium, which is of extremely significance. 
First, a pseudorapidity dependent hydrodynamic
response allows one to detect
the fluid locally, so that the transport properties of 
the QGP medium can be examined differentially. 
One such example is 
the attempt to extract a temperature dependent ratio of shear viscosity over the entropy density,
$\eta/s(T)$~\cite{Denicol:2015nhu}.
Besides, if the local gradients in pseudorapidity are accessible, the convergence behavior of the hydrodynamic
gradient expansion can be 
analyzed in realistic QGP medium with non-trivial
3+1 dimensional expansions, which would substantially extends the recent developments
with respect to systems undergoing one dimensional Bjorken expansion (cf.~\cite{Heller:2015dha,Heller:2013fn,Basar:2015ava}).

\section{Formulation} 
We first present a step-by-step derivation of the generalization of \Eq{eq:linear}.
The only assumption we have is the fluidity of QGP, so
that long wave-length (small wave-number) modes dominate. 

In heavy-ion collisions, from each collision event the single-particle distribution 
that describes the probability of particle emission
with azimuthal angle ($\phi={\rm atan}(p_y/p_x)$) and pseudorapidity 
($\zeta={\rm atanh}(p_z/|\bf p|)$) dependence can be decomposed
into Fourier modes
\be
\label{eq:prob}
P(\phi,\zeta) = \frac{1}{2\pi} \sum_{n=-\infty}^{n=+\infty} V_n(\zeta) e^{-in\phi}\,,
\ee
where the Fourier coefficient defines the complex harmonic flow
$V_n(\zeta) \equiv v_n(\zeta) e^{in\Psi_n(\zeta)}$. It is worth mentioning that the definition of flow harmonics in 
\Eq{eq:prob} differs from the usual one by a factor of normalized multiplicity distribution. 
Given the definition in \Eq{eq:prob}, the integrated 
flow coefficient is then $
V_n = \int_{-\infty}^{\infty} d\zeta \;V_n(\zeta)\,.
$
Throughout this letter, we shall focus on the second harmonic, namely, elliptic flow $V_2(\zeta)$, while
generalizations to other flow harmonics are straightforward.

Initial eccentricity $\E_2$ 
is a complex quantity, which characterizes the elliptical asymmetry of the 
three-dimensional initial density profile $\rho(\vec x_\perp, \xi)$.
Note that $\xi=\atanh(z/t)$ is the space-time rapidity, to be distinguished from the pseudorapidity $\zeta$.
In order to generalize the response relation along the longitudinal direction, 
we consider the definition for a $\xi$-dependent eccentricity
as 
\be
\label{eq:eps1}
\E_2(\xi) = -\frac{\int d^2 \vec x_\perp \; \rho(\vec x_\perp,\xi) (x+iy)^2}
{\int d\xi d^2 \vec x_\perp \; \rho(\vec x_\perp,\xi) |x+iy|^2}\,,
\ee
so that the azimuthal deformation of the density profile at
a specified space-time rapidity $\xi$ is quantified. 
\Eq{eq:eps1} is \emph{not} a unique definition that introduces dependence on
space-time rapidity in eccentricity~\cite{Pang:2015zrq,Franco:2019ihq},
but \Eq{eq:eps1} satisfies the condition:
$
\E_2 = \int_{-\infty}^{\infty} d\xi\; \E_2(\xi)\,,
$
and more importantly,
\be
\label{eq:ana}
d^{m}\E_2(\xi)/d\xi^{m}\rightarrow 0, \quad{\rm when}\quad |\xi|\rightarrow \infty\,,
\ee 
for $m=0,1,2,\ldots$,
which will be found useful.

Correspondingly, we generalize the linear 
response relation \Eq{eq:linear} 
as~\footnote{
Hereafter, without specification, 
integration over the space-time rapidity $\xi$ or the pseudorapidity $\zeta$ is taken
by default from $-\infty$ to $\infty$.
}
\be
\label{eq:resp}
V_2(\zeta) = \int d\xi \; G(\zeta-\xi) \E_2(\xi)\,,
\ee
where the real constant response coefficient is replaced by a
response function $G(\zeta-\xi)$. This function is expected entirely determined by the
fluid properties of the QGP medium, which does not fluctuate
in one specified centrality class where the system multiplicity
is a fixed constant.
\Eq{eq:resp} is a non-local and differential response relation, relating 
the eccentricity at one space-time rapidity $\xi$ at initial time to the flow observed at 
pseudorapidity $\zeta$ through the evolution of hydrodynamic modes. Note that the
form of the response function implies a boost invariant background, 
on top of  which the apparent breaking of
the boost invariant symmetry in realistic heavy-ion collisions can be accounted for as perturbations
introduced by the decomposed modes in $\E_2(\xi)$.

To identify the response function, it is
advantageous to work with a Fourier transformation, namely, for the elliptic flow
$\t V_2(k) = \int_{-\infty}^\infty d\zeta \;V_2(\zeta) e^{-ik\zeta}$ and 
initial eccentricity $\t \E_2(k) =\int_{-\infty}^\infty d\xi \;\E_2(\xi) e^{-ik\xi}$.
In terms of the wave-number $k$, \Eq{eq:resp} becomes,
\be
\label{eq:lineark}
\t V_2(k)=\t G(k) \t \E_2(k)\,.
\ee
In the long wave-length limit with $|k|\ll k^*$, corresponding to the hydrodynamic regime,
one is allowed to expand the response function in series of $k$. For instance, up to the second order
the expansion is
\be
\label{eq:expk}
\t G(k)= G_0 + ikG_1 - k^2 G_2 + O(k^3)\,,
\ee
which amounts to 
\be
\label{eq:vn_eta}
V_2(\zeta) = G_0 \E_2(\zeta) + G_1 \frac{d\E_2(\zeta)}{d\zeta} + G_2 \frac{d^2 \E_2(\zeta)}{d \zeta^2}
+ O\left(\frac{d^3}{d\zeta^3}\right)\,,
\ee
in the $\zeta$-space.
By construction, these expansion coefficients are real constants, to be determined later.
Because of the boost-invariant background, reflection symmetry is implied so that
odd order response coefficients must vanish.

Several comments are in order. First, 
the leading order term can be identified as the response of the global elliptic asymmetry, while
higher order terms ($O(k)$ and beyond) do not contribute after an integration over pseudorapidity.
Therefore, \Eq{eq:linear} will be recovered from \Eq{eq:vn_eta} after an integration over $\zeta$.
Secondly, these higher order terms in the series expansion correspond to the hydro gradients, which
characterize modes satisfying a gapless dispersion relation. As is evident in the response relation, 
these hydrodynamic gradients are specified as derivatives in space-time rapidity, as a consequence of perturbations of 
the initial state density profile along $\xi$. Thirdly, validity of the hydrodynamic gradient expansion
is determined by its convergence behavior. The radius of convergence $k^*$, for instance, is 
finite if the hydrodynamic gradient expansion is convergent~\cite{Grozdanov:2019kge}.
If, on the other hand, the gradient expansion is asymptotic, practical applications of
the response formulation 
would rely on a Borel resummation method~\cite{Heller:2015dha}. 
As will be clear in \app{sec:appA}, the existence of a finite $k^*$ results in an effective regularization of the
eccentricity distribution along space-time rapidity, which smooths out local structures.
More detailed discussion on the convergence behavior and $k^*$ will be given in Sec.~\ref{sec:secIIIB}. 

To access these expansion coefficients $G_n$'s, 
it is natural to define new sets of flow observables and
initial eccentricity variables weighted with powers of $\zeta$ (or $\xi$),
\begin{subequations}
\label{eq:vnn}
\begin{align}
V^{(n)}_2\equiv&\int d\zeta \zeta^n V_2(\zeta)/n!\,,\\
\E^{(n)}_2\equiv& \int d\xi \xi^n \E_2(\xi)/n!\,.
\end{align}
\end{subequations}
These new variables are defined according to the series expansion form \Eq{eq:vn_eta},
reflecting the nature of hydrodynamic mode evolution. Especially, $\E_2^{(n)}$ provides
an alternative mode decomposition of the initial eccentricity distribution along
the space-time rapidity, comparing to those developed on grounds of initial state 
geometrical fluctuations~\cite{Bzdak:2012tp,Jia:2015jga}.
Using integration by parts repeatedly, the generalized linear response relation \Eq{eq:resp} can be written
in an alternate form as 
\be
\label{eq:linear_eta}
V_2^{(n)}=\sum_{i=0}^n (-1)^i G_i \E_2^{(n-i)}\,.
\ee
Note that \Eq{eq:ana} has been taken into account implicitly to obtain \Eq{eq:linear_eta}. Note also 
that the leading order relation $V_2^{(0)}=G_0 \E_2^{(0)}$ is the linear response 
relation presented in \Eq{eq:linear},
and the response coefficient can be calculated in event-by-event hydrodynamic simulations~\cite{Noronha-Hostler:2015dbi}:
$
G_0 = \bra V_2^{(0)} \E_2^{(0)*}\ket\Big/\bra \E_2^{(0)}\E_2^{(0)*}\ket
$, where the angular brackets indicate event average.
Given the leading order response coefficient $G_0$, one realizes a linear response
relation between $\E_2^{(0)}$ and $V_{2}^{(1)}-G_0 \E_2^{(1)}$, with the slope identified as
$G_1$.  
In a similar way,  a set of linear relations can be realized
between $\E_2^{(0)}$ and $V_2^{(n)}-\sum_{i=0}^{n-1} (-1)^{i}G_i \E_2^{(n-i)}$, with
$G_n$ calculated recursively as
\begin{align}
\label{eq:eva_Gn}
&(-1)^n
G_n = \cr 
&\bra\left(V_2^{(n)}-\sum_{i=0}^{n-1} (-1)^{i}G_i \E_2^{(n-i)}\right)\E_2^{(0)*} \ket\Big/\bra \E_2^{(0)}\E_2^{(0)*}\ket\cr
\end{align}

\section{Event-by-event hydrodynamic simulations}

\begin{figure}
\begin{center}
\includegraphics[width=0.4\textwidth] {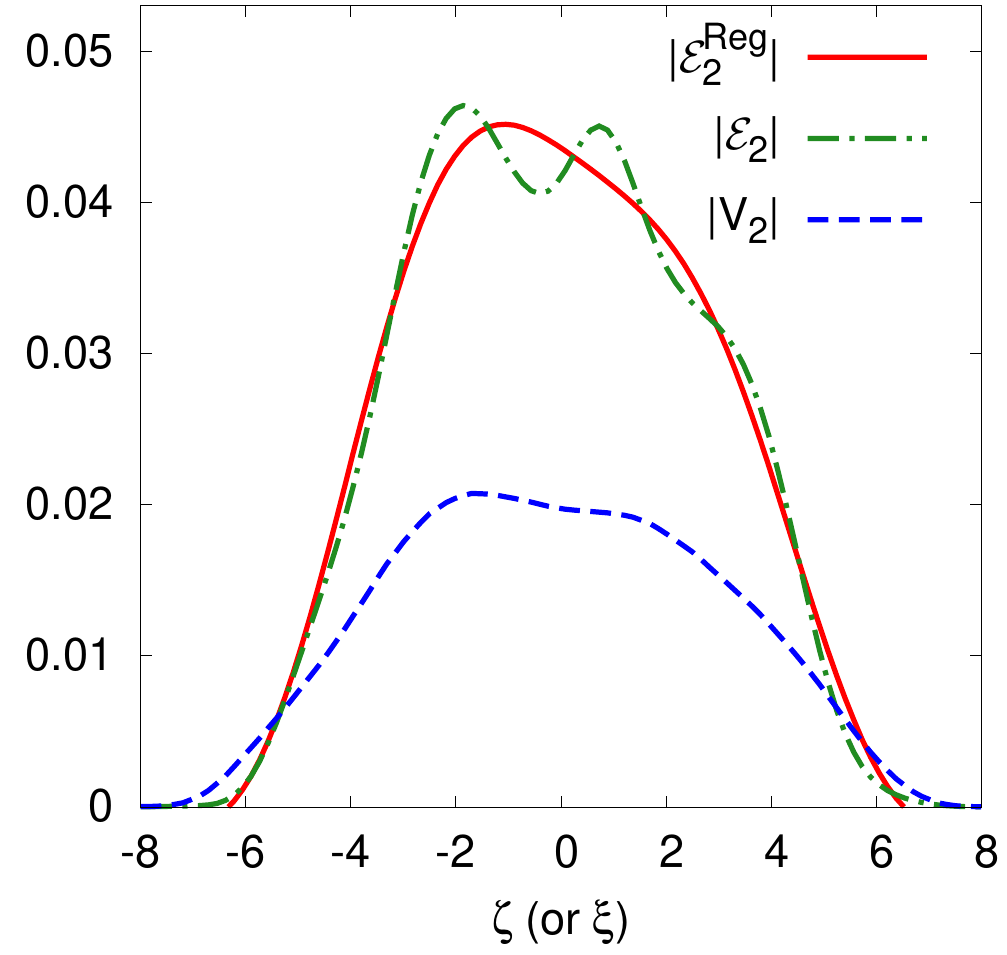}
\caption{ (Color online) 
Longitudinal distribution of initial eccentricity and flow in one collision event from 
numerical hydrodynamic simulations. The green dash-dotted line and the blue dashed lines
correspond to the magnitude of eccentricity and elliptic flow, respectively. With $k^*=1.2$, 
the red solid line shows the regularization of initial eccentricity distribution due to 
the finite radius of convergence 
of the longitudinal hydrodynamic response, according to \Eq{eq:reg}.
 \label{fig:oneEve}
}
\end{center}
\end{figure}

To test the analytical formulations of the longitudinal hydrodynamics response,
we carry out realistic event-by-event hydrodynamic simulations,
for
the Pb-Pb collision system with $\sqrt{s_{NN}}=2.76$ TeV at the LHC, for events in centrality class of 30-40\%.
This is realized by the 3+1D MUSIC~\cite{Schenke:2010nt,Schenke:2010rr,Paquet:2015lta}, with respect to random 3D initial conditions generated by the 
AMPT model~\cite{Lin:2004en,Lin:2014tya}. The initial density profile is obtained in a similar strategy as 
in Ref.~\cite{Li:2017slc}, with an overall constant factor adjusted
to reconcile the mismatch of viscosity in AMPT and viscous hydrodynamics~\cite{Pang:2015zrq,Chattopadhyay:2017bjs}. Note that non-trivial longitudinal density distribution with fluctuations in the AMPT model have been implemented. 
In \Fig{fig:oneEve}, the distribution of eccentricity
magnitude along
space-time rapidity in one collision event is shown as the green dash-dotted line, where bumps appear
as a consequence of longitudinal fluctuations.

\begin{figure}
\begin{center}
\includegraphics[width=0.4\textwidth] {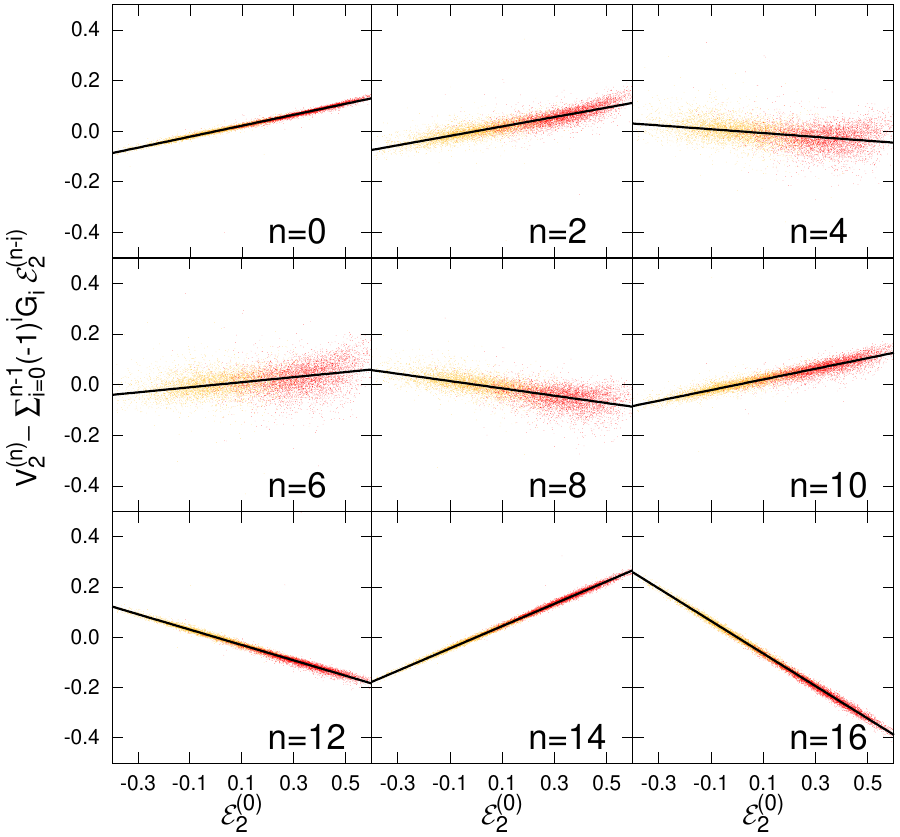}
\caption{ (Color online) Scatter plot showing the linear 
relations
between $\E_2^{(0)}$ and $V_2^{(n)}-\sum_{i=0}^{n-1} (-1)^{i}G_i \E_2^{(n-i)}$, with coefficients $G_n$'s solved recursively with
respect to \Eq{eq:eva_Gn} (slopes of solid lines), 
using event-by-event 3+1D hydro simulations with $\eta/s=0.08$. 
Red and yellow points are for
the real and imaginary parts in the linear relation respectively.
Slopes of
odd orders (not shown) are consistent with zero.
 \label{fig:linear}
}
\end{center}
\end{figure}

MUSIC solves the second order viscous hydrodynamics 
with respect to a 3+1D expanding system, in which a cross-over from the QGP to hadron gas is
implied in an equation of state obtained from lattice QCD. We calculate elliptic flow $V_2$ from thermal pions.
Throughout our analyses, statistical errors from simulations
are estimated using the Jackknife resampling method.
In the current study, we only consider shear viscous corrections
with a constant ratio of shear viscosity to the entropy density, $\eta/s$,
while second order transport coefficients are fixed accordingly.
More details on the numerical simulations of MUSIC, such as the equation of state, the freeze-out prescription, etc.,  
can be found in Ref.~\cite{Schenke:2010nt,Schenke:2010rr,Paquet:2015lta} and references therein.

Given the results from hydrodynamic simulations, the constant response coefficients can be calculated, 
as well as the linear relations
between $\E_2^{(0)}$ and $V_2^{(n)}-\sum_{i=0}^{n-1} (-1)^{i}G_i \E_2^{(n-i)}$.
Scatter plots in \Fig{fig:linear} present these linear relations 
respectively from the real part (red points) and the imaginary part (yellow points),
from event-by-event hydro simulations of approximately 5000 events with a constant $\eta/s=0.08$.
For even orders up to $n=16$, apparent linear relations are shown, with slopes compatible with
those calculated with respect to \Eq{eq:eva_Gn}.~\footnote{
For higher order $n$, contributions from the edge of the longitudinal distributions become more
significant, which lead to numerical uncertainties of the calculated variables, 
such as $V_2^{(n)}$. Nonetheless, we have checked that such uncertainties are sufficiently suppressed 
with respect to the current numerical
settings, with $\xi\in [-8,8]$ and $\zeta\in [-8,8]$. 
}
We notice that slopes of odd orders are consistent with zero within errors, as anticipated. 

\subsection{Two-point correlation of $V_2$} 

The pseudorapidity dependent response relation \Eq{eq:resp} 
is non-local, which implies that the generation of
$V_2$ at one pseudorapidity receives contributions from other 
space-time rapidities.  But the response along longitudinal direction
is limited by the speed of sound in fluid systems. This effect
can be shown in the analysis of two-point correlations.
We define
\be
\label{eq:width_v2}
\bra (\Delta \zeta)^2\ket
\equiv
\frac{\int d\zeta d \zeta' \bra V_2(\zeta) V_2^*(\zeta')\ket(\zeta'-\zeta)^2 }
{\int d\zeta d\zeta' \bra V_2(\zeta) V_2^*(\zeta')\ket},
\ee
to characterize the length of the two-point correlation measured via elliptic flow
at different pseudo-rapidities. 
With the response relation derived in \Eq{eq:vnn} and \Eq{eq:linear_eta}, in particular, 
considering the fact that $G_1=0$, it can be proved that 
\be
\label{eq:width_g2g0}
\bra (\Delta \zeta)^2\ket=\bra (\Delta \xi)^2\ket + 4G_2/G_0\,.
\ee
The length of the initial state eccentricity two-point correlation  
$\bra (\Delta \xi)^2\ket$ is defined according to \Eq{eq:width_v2} through 
$\E_2(\xi)$.
\Eq{eq:width_g2g0} states that the increase of the two-point correlation length in elliptic flow
comparing to that in the initial eccentricity is purely an effect of fluid dynamics.
This can be understood as a direct consequence of sound propagation, reflected in the ratio
$G_2/G_0$, that initial 
perturbations propagate in speed of sound $c_s$ along the longitudinal direction~\cite{Kapusta:2011gt}.
Since sound propagation is damped as a result of fluid dissipations: Shear, bulk,
etc., in dissipative fluids the two-point correlation at the final stage is reduced.

\begin{figure}
\begin{center}
\includegraphics[width=0.4\textwidth] {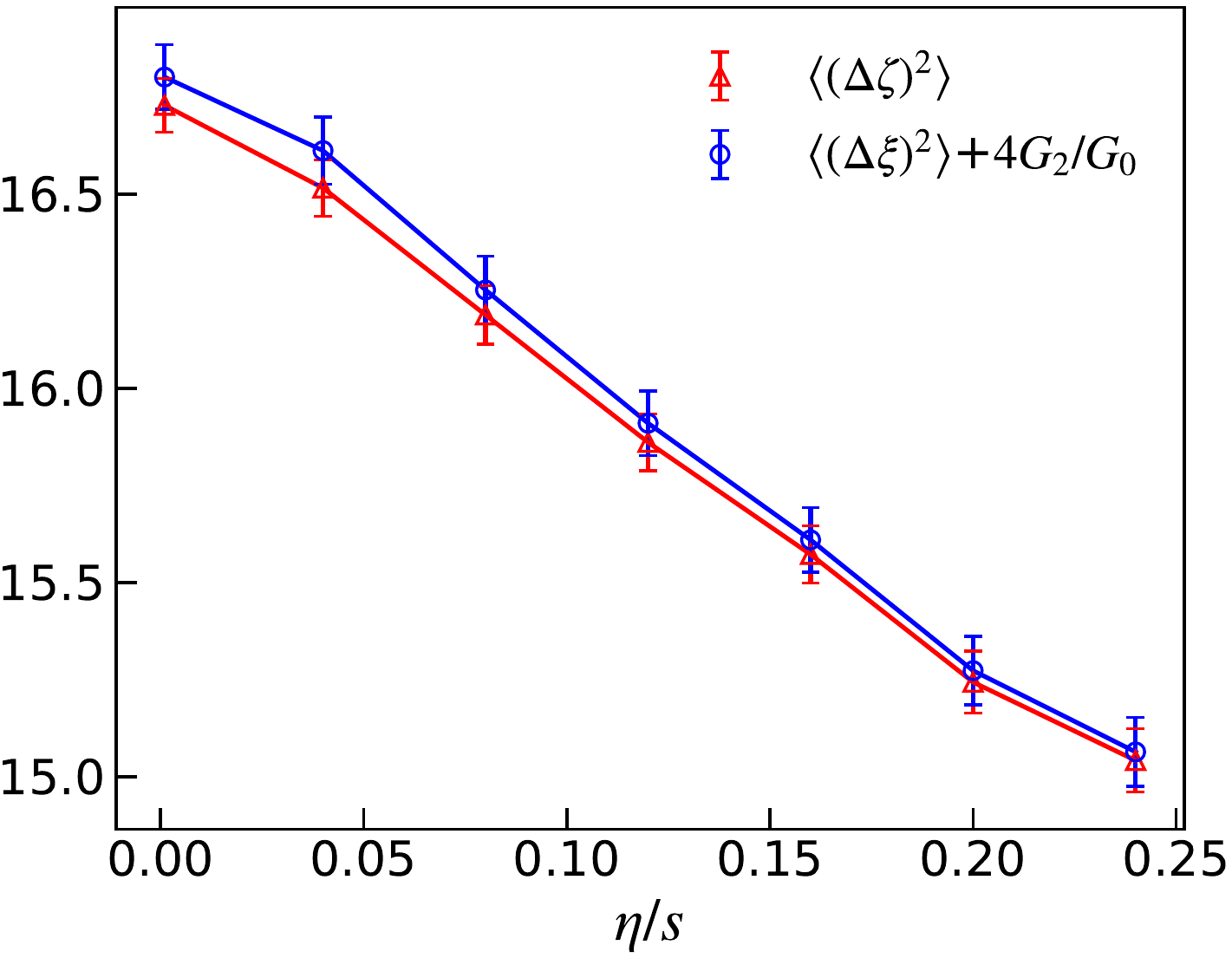}
\caption{(Color online) Two-point correlation length of $V_2$ and the 
expectations from \Eq{eq:width_g2g0}, as a function of $\eta/s$,
for the Pb-Pb system of $\sqrt{s_{NN}}=2.76$ TeV in centrality class 30-40\%, 
where the initial two-point correlation length is $\bra (\Delta \xi)^2\ket=12.81 \pm 0.08$.
 \label{fig:width}
}
\end{center}
\end{figure}

\Fig{fig:width} shows the numerical results of about 1000 events from hydrodynamic simulations with different 
values of $\eta/s$.  The obtained length of the two-point correlation of $V_2$ is plotted as a function
of $\eta/s$, which agrees with expectation from 
\Eq{eq:width_g2g0} within statistical errors. When $\eta/s\to 0$, the length of two-point correlation
approaches an upper bound determined by the sound horizon of QGP medium.
Reduction of the correlation length, or the ratio $G_2/G_0$, due to shear viscous correction is clear
and systematic.
Subject to boundary corrections due to the experimental acceptance
along pseudorapidity, $\bra (\Delta \zeta)^2\ket$ is a measurable in heavy-ion
experiments, providing a novel probe of 
$\eta/s$ in the QGP medium.

\subsection{The radius of convergence} \label{sec:secIIIB}

\begin{figure}
\begin{center}
\includegraphics[width=0.4\textwidth] {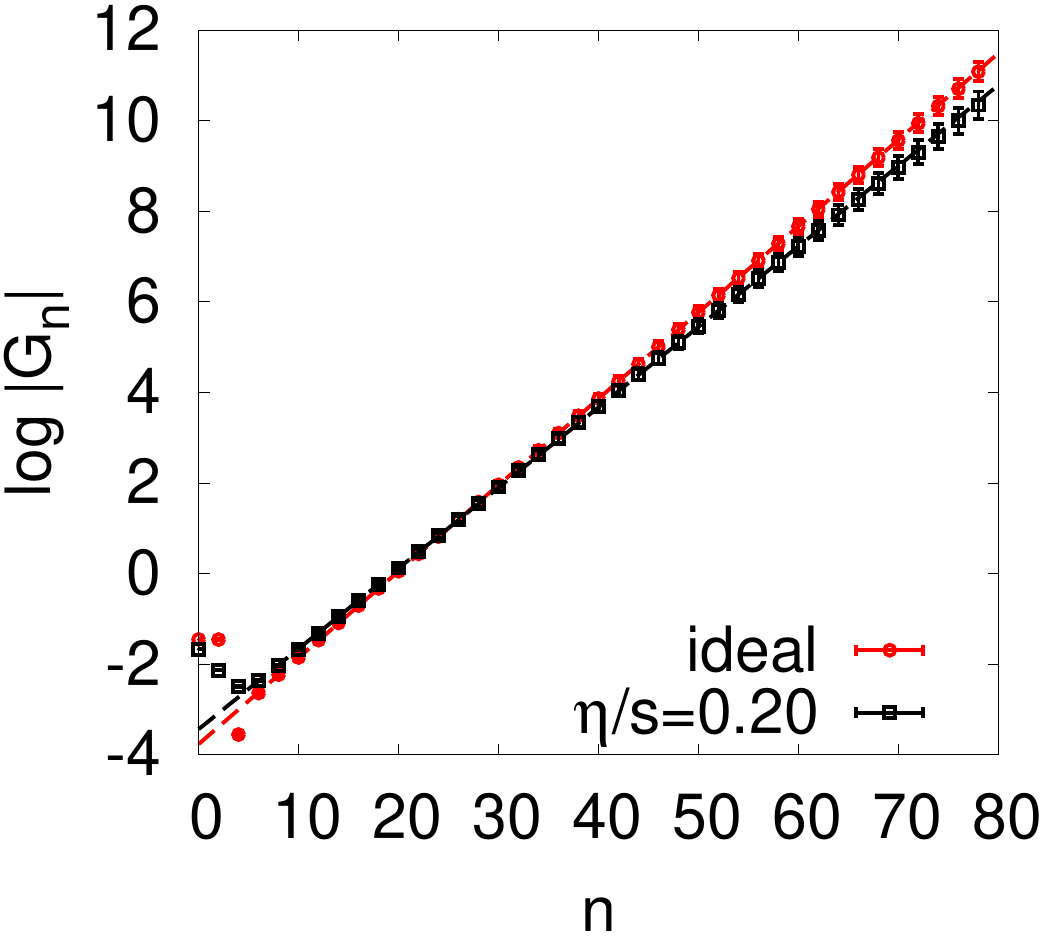}
\caption{(Color online) Exponential increase of the expansion coefficients $|G_n|$ with respect to $n$. Lines are linear fit with slope 0.190 (ideal) and 0.176 ($\eta/s=0.2$).
 \label{fig:gns} 
}
\end{center}
\end{figure}

\begin{figure}
\begin{center}
\includegraphics[width=0.4\textwidth] {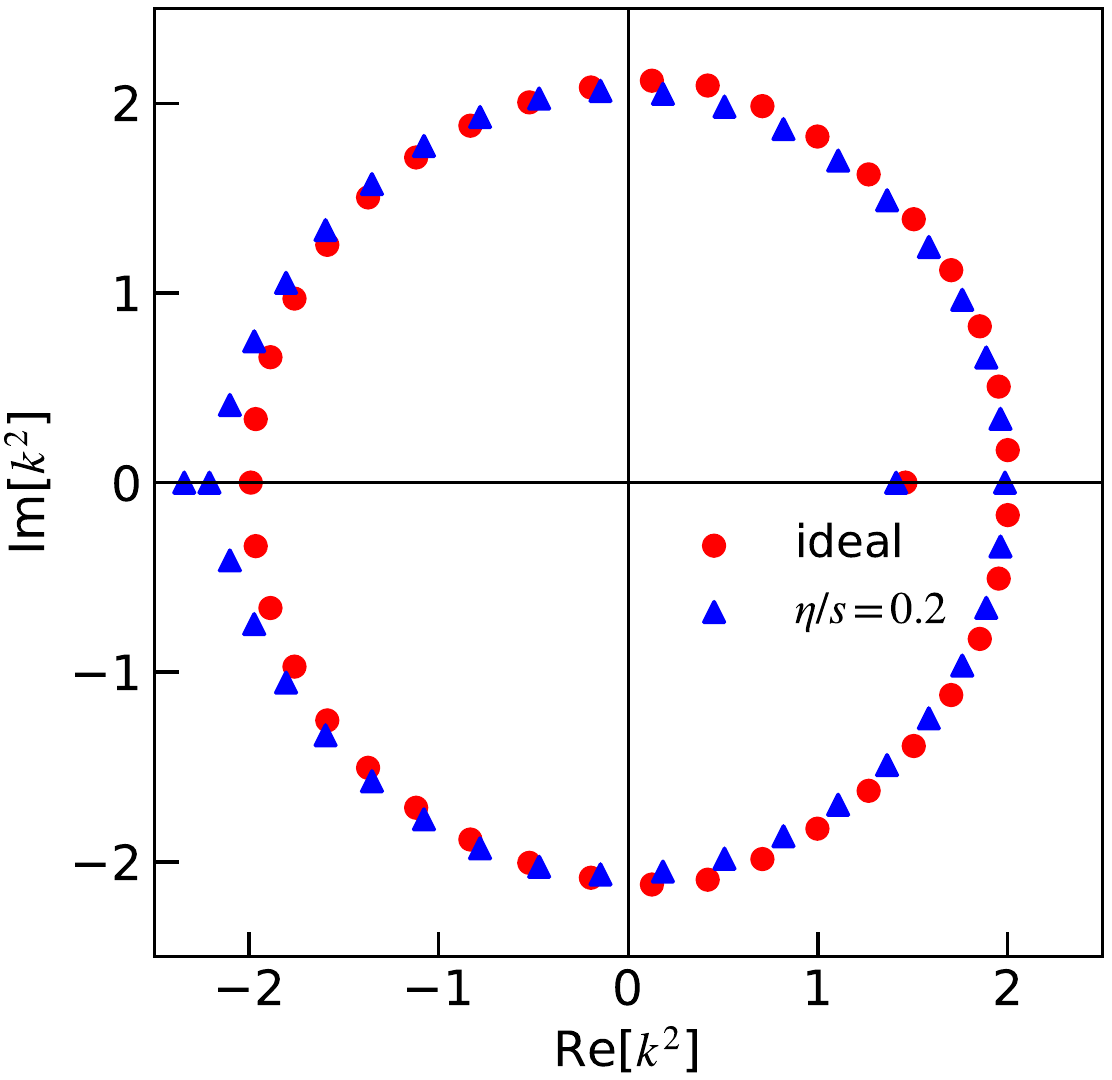}
\caption{(Color online) Pole structure in the complex $k^2$-plane from a Pad\'e approximation of 
the gradient expansion. The pole closest to origin is on the positive real axis:
$(1.4615,0)$ for the ideal case and $(1.4126,0)$ for the case of $\eta/s=0.2$.
 \label{fig:pade}
}
\end{center}
\end{figure}

With the expansion coefficients calculated from 
event-by-event hydrodynamic simulations, 
we are able to explore the
convergence behavior of the hydrodynamic gradient expansion in
\Eq{eq:expk}. 
Especially, the gradient expansion corresponds to the fluid properties in 
a \emph{realistic}, 3+1D expanding QGP medium obeying lattice equation of state. 

Owing to the reflection
symmetry, we ignore contributions from odd orders of the expansion. For $n>2$, 
we notice that the sign of $G_n$ of even orders flips. In \Fig{fig:gns}, the calculated even
order coefficients are shown, from hydrodynamic simulations of roughly 1000
events, using $\eta/s=0.001$ (ideal) and $\eta/s=0.2$, respectively. 
For $n>4$, these $G_n$'s grow exponentially: $\log|G_n|\propto \alpha n$, 
which implies a \emph{finite} radius of convergence. 
Therefore, the hydrodynamic gradient expansion in this current case is \emph{convergent}.
The slope in \Fig{fig:gns} determines the radius
of convergence. 
Alternatively, one may also apply a diagonal Pad\'e approximation for the 
gradient expansion. 
Convergence behavior can thus be revealed from the singularity 
structure of the Pad\'e approximation.
The resulted poles in the complex $k^2$-plane are shown in \Fig{fig:pade}. 
When $n>70$, the singularity 
structure of the 
Pad\'e approximation stabilizes. In particular, in addition to those poles lying around a circle, 
there exists one pole closest to the origin on the positive real axis, which determines the radius of 
convergence.  

From both ways, 
the obtained radii of convergence are finite. Numerical values of the radii are 
consistent, which we identify as the upper bound of the gradient expansion $k^*$. 
From ideal hydrodynamic simulations,
we find $k^*=1.209$ (slope) and $1.209$ (Pad\'e), 
while for the case of $\eta/s=0.2$ we find $k^*=1.192$ (slope) and $1.189$ (Pad\'e).
When shear viscosity decreases, there seems to be a trend that $k^*$ 
grows~\cite{Grozdanov:2019kge,Grozdanov:2019uhi}. However, due to the effect of expansion,
we find that $k^*$ saturates to a constant in an ideal fluid even though the mean free path 
approaches zero.

For a medium system away from local equilibrium, it is known that the gradient expansion consisting of higher
order dissipative corrections is asymptotic~\cite{Heller:2015dha,Heller:2013fn,Basar:2015ava}. Origin of such divergence can be intuitively attribute to the fact
that the number of higher order viscous terms has a factorial increase. However, it is different in the current case 
concerning perturbations on top of the medium system in or close to local equilibrium, where 
gradients are taken into account with respect to the dispersion relation~\cite{Withers:2018srf,Grozdanov:2019kge,Grozdanov:2019uhi}.

In the current study, the existence of a \emph{finite} radius of convergence of the gradient expansion  
has an important physical interpretation. The upper bound $k^*$ of the expansion implies a 
length scale limit in the space-time rapidity of the initial eccentricity profile $\E_2(\xi)$. This is the scale
that determines the resolution of fluid response to initial state local structures. 
Although for a static medium 
this effect can be generically understood
as the fact that the fluid behavior of a medium system must be visible at a scale greater than the mean
free path and one expects $k^*\sim 1/l_{\rm mfp}$, in the medium with non-trivial 3+1D expansion it becomes more complicated.
As a consequence of the hydro response and finite $k^*$, a 
regularization scheme for the initial eccentricity profile can be introduced, leading to a regularized eccentricity profile
$\E_2^{\rm Reg}(\xi)$.  Derivation of the regularization scheme 
is given in \app{sec:appA}.
In \Fig{fig:oneEve}, one finds in the resulted $\E_2^{\rm Reg}$ the local 
perturbations are smoothed out, giving rise to a consistent shape comparing with the distribution of
flow where bumpy structures are merely seen.

\section{Summary and discussions}
We have derived the generalized formulation of pseudorapidity dependent hydrodynamic response. 
The generalized form is a non-local relation that interprets the formation of elliptic
flow at different pseudo-rapdities as a consequence of longitudinal fluid response, 
which provides a novel picture of longitudinal flow generation that differs from 
some previous studies (cf. Ref.~\cite{Bozek:2015bha,Sakai:2018sxp}).
The formulation is 
applicable to other flow harmonics as well, when complexities involving nonlinearities do not
arise~\cite{Noronha-Hostler:2015dbi,Teaney:2012ke}.

The longitudinal fluid response can be quantified by a set of constant coefficients $G_n$'s, and a new set of 
flow observables in \Eq{eq:vnn}. This is the main result of
this letter, which has been shown valid through realistic event-by-event hydrodynamic simulations
for a 3+1D expanding system in heavy-ion collisions. The second order expansion coefficient $G_2$ 
is found to be sensitive to the sound mode propagation, contributing to the increase of
two-point correlation length of the elliptic flow. Shear viscous correction in the fluid reduces two-point
correlation, as expected from the damping of sound. Therefore, two-point correlation of flow 
provides a new measurable to investigate the 
shear viscosity of the QGP medium in experiments.

Within the formulation of 
pseudorapidity dependent hydrodynamic response, 
the expansion coefficients $G_n$'s can be calculated with respect to 
a realistic expanding QGP medium.
Given these expansion coefficients, hydrodynamic gradient expansion can be analyzed.
In this current study, we solved 3+1D MUSIC with respect to fluctuating initial conditions, which
represents the fluid response to perturbations on top of a fluid medium that 
obeys second order viscous hydrodynamics.
Generation of the elliptic flow from gradients along the longitudinal direction thus reflects the 
evolution of sound and shear diffusive modes, as in the hydrodynamic dispersion relations.
A finite radius of convergence is observed, which confirms the finding that hydrodynamic
gradient expansion associated with the dispersion relation is 
convergent~\cite{Withers:2018srf,Grozdanov:2019kge,Grozdanov:2019uhi}.

\emph{Acknowledgements} -- 
We thank Subikash Choudhury for his contributions at an early stage of this
project. We thank Xiao-Liang Xia for his help in generating the hydrodynamic initial conditions.
We thank Huan Zhong Huang for carefully reading the manuscript and for his very
inspiring comments. We thank Jean-Yves Ollitrault for pointing out the fact that odd order expansion
coefficients must vanish with respect to the boost invariant background. 
This work is supported by National Natural Science Foundation of China (NSFC) 
under Grant No. 11975079 and the China Postdoctoral Science Foundation under Grant No. 2019M661333 (HL).

\appendix

\section{Regularization of the eccentricity distribution with finite $k^*$}
\label{sec:appA}

The existence of a finite radius of convergence $k^*$ in the hydrodynamic gradient expansion 
effectively introduces regularization of the initial state eccentricity distribution. This effect can
be formulated through corrections to the original Fourier transform of the eccentricity distribution, 
\begin{align}
\label{eq:reg}
\E_2^{\rm Reg}(\xi) =& \int_{-k^*}^{k^*} \frac{d k}{2\pi} e^{ik\xi}\t \E_2(k) 
 =\int_{-k^*}^{k^*} \frac{d k}{2\pi} e^{ik\xi} \int_{-\infty}^\infty d\xi' \E_2(\xi') e^{-ik \xi'}\cr
=& \int_{-\infty}^\infty d\xi' \E_2(\xi') R(\xi-\xi'; k^*)\,,
\end{align}
where accordingly a regularization function is defined,
\be
\label{eq:Rfunc}
R(\xi-\xi';k^*) \equiv \frac{\sin(k^*(\xi-\xi'))}{\pi (\xi-\xi')}
=\int_{-k^*}^{k^*} \frac{dk}{2\pi} \; e^{ik(\xi-\xi')}\,.
\ee
Apparently, for $k^*\to \infty$, the function reduces to a Dirac delta function: 
$R(\xi-\xi'; k^*)\to \delta(\xi-\xi')$. In the opposite limit: $k^*\to 0$, one finds that
\be
\frac{d}{d \xi} R(\xi-\xi';k^*) \to 0\,, 
\ee
which gives rise to a boost invariant $\E_2^{\rm Reg}(\xi)$.

\bibliography{references}
\end{document}